\documentclass[aps]{revtex4}

\begin{document}

\title{Comment on ''Superconducting decay length in a ferromagnetic metal''\\
}
\author{A.F. Volkov$^{1,2}$, F.S. Bergeret$^{3}$ and K.B.Efetov$^{1,4}$}
\affiliation{ 
$^{(1)}$Theoretische Physik III,\\
Ruhr-Universit\"{a}t Bochum, D-44780 Bochum, Germany\\
$^{(2)}$Institute of Radioengineering and Electronics of the Russian Academy%
\\
of Sciences, 103907 Moscow, Russia\\
$^{(3)}$ Departamento de F\'{i}sica Te\'{orica} de la Materia
Condensada C-V, Universidad Aut\'{o}noma de Madrid, E-28049
Madrid, Spain\\
$^{(4)}$L.D. Landau Institute for Theoretical Physics, 117940 Moscow, Russia%
\\
}
\begin{abstract}
In the paper \textquotedblright Superconducting decay length in a
ferromagnetic metal\textquotedblright\ by Gusakova, Kupriyanov and Golubov
[Pis'ma v ZhETF 83, 487 (2006); cond-mat/0605137], the authors claim that
they solved the linearized Eilenberger equation in the ferromagnetic region
of an S/F heterostructure at arbitrary mean free path. In this comment we
show that the solution suggested by the authors is not correct and explain
details of the \emph{exact }solution found by us in an earlier work several
years ago (Ref.[2]).
\end{abstract}

\maketitle

\section{Introduction}

\bigskip

\bigskip In a recent paper \cite{Kupriyanov} Gusakova, Kupriyanov and
Golubov studied theoretically the decay length over which the condensate
function $f$ decreases in the ferromagnetic region of a S/F structure (S and
F denote a superconductor and ferromagnet). In order to determine the
decaying length they have solve the linearized Eilenberger equation
representing the solution in the form 
\begin{equation}
\ f(x)=C(\theta )\exp (-x/\xi )\;,  \label{1}
\end{equation}
where $\theta $ \ is an angle between the direction of the momentum and the $%
x$-axis (the x-axis is perpendicular to the S/F interface). In this
solution the function $C(\theta )$ depends only on $\theta $ and the decay
length $\xi $ is assumed to be independent of this angle. Substituting this
Ansatz into the linearized Eilenberger equation the authors obtain
expressions for ${C(\theta )}$ and the decay length $\xi $. They analyze the
dependence of the decay length on the exchange field $h$ and the mean free
path for an arbitrary magnitude of the product $h\tau $ (we set the Planck
constant equal to 1).

In this Comment we would like to point out that:

1. The solution of the form (\ref{1}) for the linearized Eilenberger
equation is not correct and may serve only for rough estimates for the decay
length $\xi $ in a particular limit.

2. The linearized Eilenberger equation can be solved exactly for an
arbitrary value of product $h\tau .$ The exact solution has been obtained in
Ref. \cite{BVE} for arbitrary values of $h\tau $ (contrary to the statement
of the authors of Ref.\cite{Kupriyanov} that the solution was obtained in 
\cite{BVE} in the clean limit only) and used later in subsequent theoretical 
\cite{Fominov,Kharitonov} and experimental studies \cite{Blum,Palevski}.

In Ref.\cite{BVE} the solution of the linearized Eilenberger equation for
the Josephson S/F/S junction is presented explicitly. In order to avoid a
literal repetition of that result we use the method of Ref. \cite{BVE}, to
obtain the exact solution for a simpler S/F structure considered in \cite
{Kupriyanov}, thus directly comparing the result with ''the solution''\ of
Eq. (\ref{1}) from that work. From a general formula we obtain simple
expressions for $f(x)$ in the limiting cases of small and large values of
the product $h\tau $ (the diffusive and quasi-ballistic cases
correspondingly). {To make the analysis more transparent, we neglect
the spin-orbit interaction.}

The solution of the linearized Eilenberger equation $\hat{f}(x)$ can be
represented as a sum of a symmetric $\hat{s}(x)$ and antisymmetric $\hat{a}%
(x)$ parts$:\hat{f}(x)=\hat{s}(x)$ $+$ $\hat{a}(x)$. The functions $\hat{f}%
(x),$ $\hat{s}(x)$ and $\hat{a}(x)$ are matrices in the particle-hole space.
The antisymmetric part $\hat{a}(x)$ is related to the symmetric one via the
formula

\begin{equation}
\hat{a}=-\mathrm{sgn}\omega \left( \mu l/\kappa _{\omega }\right) \hat{\tau}%
_{3}\partial _{x}\hat{s}  \label{2}
\end{equation}
and the symmetric part $\hat{s}(x)$ in the ferromagnetic region ($x>0$)
obeys the equation \cite{BVE}

\begin{equation}
\mu ^{2}l^{2}\partial _{xx}^{2}\hat{s}-\kappa _{\omega }^{2}\hat{s}=-\kappa
_{\omega }\left\langle \hat{s}\right\rangle \;  \label{3}
\end{equation}
where $l=v_{F}\tau $ is the mean free path, $\mu =\cos \theta ,$ $\kappa
_{\omega }=(1+2|\omega _{m}|\tau )-\mathrm{sgn}\omega 2ih\tau ,$ $\omega
_{m}=\pi T(2m+1)$ is the Matsubara frequency. The angle brackets mean the
angle averaging

\begin{equation}
\left\langle \hat{s}\right\rangle \;=(1/2)\int_{-1}^{1}d\mu \hat{s}(\mu )
\label{4}
\end{equation}
Eqs (\ref{2} - \ref{3}) are complemented by the boundary condition \cite
{Zaitsev}

\begin{equation}
\hat{a}\mid _{x=0+}=\gamma (\mu )\mathrm{sgn}\omega \left( \hat{\tau}_{3}%
\hat{f}_{s}\right) \;,  \label{5}
\end{equation}
where $\gamma (\mu )=T(\mu )/4$ and $T(\mu )$ is the transmission
coefficient. The matrix $\hat{f}_{s}=i\hat{\tau}_{2}\Delta /\sqrt{\omega
_{m}^{2}+\Delta ^{2}}\equiv i\hat{\tau}_{2}f_{s}$ is the matrix condensate
function of the superconductor (the BCS-function). The boundary condition (%
\ref{5}) is valid in the case of a small transmission coefficient $T(\mu )$
when the linearization of the Eilenberger equation is possible.

We can formally continue the solution of Eq. (\ref{3}) into the region of
negative $x$ in a symmetric way taking into account the boundary condition (%
\ref{5}). Therefore Eq. (\ref{3}) may be written in the form

\begin{equation}
\mu ^{2}l^{2}\partial _{xx}^{2}s-\kappa _{\omega }^{2}s=-\kappa _{\omega
}[\left\langle s\right\rangle \;+2\gamma (\mu )l\mu f_{s}\delta (x)]
\label{6}
\end{equation}
where the function $s(x)$ is defined as $\hat{s}(x)=s(x)i\hat{\tau}_{2}.$
The boundary condition (\ref{5}) is satisfied automatically due to the last
term on the right-hand side. For the Fourier transform $s_{k}=\int
dxs(x)\exp (-ikx)$ we get from Eq.(\ref{6})

\begin{equation}
s_{k}(\mu )=\frac{\kappa _{\omega }}{M(\mu )}\{\langle s_{k}(\mu )\rangle
+2\gamma (\mu )\mu lf_{s}\}\; ,  \label{7}
\end{equation}
where $M(\mu )=(kl\mu )^{2}+\kappa _{\omega }^{2}.$

Averaging $s_{k}(\mu )$ over $\mu $ and substituting the result into Eq.(\ref
{7}), we obtain finally

\begin{equation}
s_{k}(\mu )=\frac{2\kappa _{\omega }lf_{s}}{M(\mu )}\{\frac{\kappa _{\omega }%
}{1-\kappa _{\omega }\langle 1/M\rangle }\langle \frac{\gamma (\mu )\mu }{M}%
\rangle +\gamma (\mu )\mu \}  \label{8}
\end{equation}

The inverse Fourier transformation

\begin{equation}
s(x,\mu )=\int \frac{dk}{2\pi }s_{k}(\mu )\exp (ikx)  \label{9}
\end{equation}
determines the exact solution for Eq.(\ref{6}) at arbitrary mean free path.
It is obvious that the function $s(x,\mu )$ cannot be represented as a
product of two functions, such that one of them would depend only on $\mu $
and the other one only on $x$.

The average $\langle 1/M\rangle $ can easily be calculated

\begin{equation}
\langle 1/M\rangle =\frac{1}{kl\kappa _{\omega }}\tan ^{-1}\frac{kl}{\kappa
_{\omega }}  \label{10}
\end{equation}

Let us consider limiting cases, where one can obtain analytical expressions
for the condensate function $s(x,\mu ).$

a) diffusive limit: $h\tau <<1.$

In this case, characteristic wave vectors $k$ are much smaller than $1/l.$
Expanding the function on the right-hand side in Eq.(\ref{10}) in powers of $%
kl$, we obtain $1-\kappa _{\omega }\langle 1/M\rangle \approx
l^{2}(k^{2}+k_{h}^{2})/3.$ The behavior of $s(x,\mu )$ at large $x$ (larger
than the mean free path $l$) is determined by the pole in the first term in
Eq.(\ref{8}){.} Calculating the residue at this pole, we get

\begin{equation}
s(x,\mu )\approx \frac{3f_{s}}{k_{h}l}\langle \gamma (\mu )\mu \rangle \exp
(-k_{h}x)  \label{11}
\end{equation}
where $k_{h}^{2}=2[|\omega _{m}|-\mathrm{sgn}\omega _{m}ih]/D,$ $D=lv/3$ is
the diffusion coefficient. This result can be easily obtained from the
Usadel equation. One can see that $s(x,\mu )$ is small and the linearization
of the Eilenberger equation is possible provided the condition $3\sqrt{3}%
\langle \gamma (\mu )\mu \rangle /\sqrt{h\tau }<<1$ is fulfilled. Only in
this case the characteristic length of the condensate decay (at large $x$)
does not depend on the angle $\theta $. At $x$ of the order of the mean free
path the solution of Eq.(\ref{6}) depends on $\theta $. Thus, a solution of
the form Eq. (\ref{1}) is only valid in the diffusive limit and for large $x$%
. However, in the case of strong ferromagnets or not too short mean free
path the solution depends on angles at any $x$, as we will show.

b) quasi-ballistic case: $h\tau >>1.$

In this case the parameter $\mid \kappa _{\omega }\mid \approx h\tau $ is
large and therefore the first term in the curly brackets in Eq. (\ref{8}) is
small. Calculating the integral in Eq.(\ref{9}), we get

\begin{equation}
s(x,\mu )\approx \gamma (\mu )f_{s}\exp (-k_{\omega }x/l\mu ),\text{ \ \ }%
\mu >0  \label{12}
\end{equation}

We see that in this case the decay length $\xi _{F}$ which is equal to $\xi
_{F}=\min \{v\mu /T,\mu l\}$ is angle dependent. The condensate function
oscillates in space with a small (compared to $\xi _{F}$) period $\xi
_{osc}=l\mu /(2\pi h\tau ).$ It is clear that the condensate function can
not be presented in the form (\ref{1}).

In a general case of arbitrary product $h\tau $ ($h\tau \approx 1$) the
spatial dependence of the Cooper pair wave function $f$ can be found by
numerical integration of Eqs.(\ref{7}-\ref{8}).


\end{document}